\documentclass[aps,pra,reprint,showpacs,twocolumn,nofootinbib]{revtex4-1}
\usepackage{graphicx}
\usepackage{amsmath}
\usepackage{amssymb}
\usepackage{amsfonts}
\usepackage{bbm}
\usepackage{bm}
\usepackage{verbatim}
\usepackage{cancel}
\usepackage{wasysym}
\usepackage[bookmarks=false,colorlinks=true,urlcolor=blue,citecolor=blue,linkcolor=blue]{hyperref}
\usepackage[normalem]{ulem}

\renewcommand\({\ensuremath \left(}
\renewcommand\){\ensuremath \right)}
\renewcommand\[{\ensuremath \left[}
\renewcommand\]{\ensuremath \right]}
\def\:={\,\raisebox{0.85pt}{.}\hspace{-2.78pt}\raisebox{2.85pt}{.}\!\!=\,}
\def\=:{\,=\!\!\raisebox{0.85pt}{.}\hspace{-2.78pt}\raisebox{2.85pt}{.}\,}

\begin{document}

\title{Weak Ergodicity Breaking and Quantum Many-Body Scars in Spin-1 XY Magnets}

\author{Michael Schecter}
\affiliation{Condensed Matter Theory Center and Joint Quantum Institute, Department of Physics, University of Maryland, College Park, Maryland 20742, USA}

\author{Thomas Iadecola}
\affiliation{Condensed Matter Theory Center and Joint Quantum Institute, Department of Physics, University of Maryland, College Park, Maryland 20742, USA}

\date{\today}

\begin{abstract}
We study the spin-1 XY model on a hypercubic lattice in $d$ dimensions and show that this well-known nonintegrable model hosts an \emph{extensive} set of anomalous finite-energy-density eigenstates with remarkable properties. Namely, they exhibit subextensive entanglement entropy and spatiotemporal long-range order, both believed to be impossible in typical highly excited eigenstates of nonintegrable quantum many-body systems.
While generic initial states are expected to thermalize, we show analytically that the eigenstates we construct lead to weak ergodicity breaking in the form of persistent oscillations of local observables following certain quantum quenches---in other words, these eigenstates provide an archetypal example of so-called ``quantum many-body scars."  This work opens the door to the analytical study of the microscopic origin, dynamical signatures, and stability of such phenomena.
\end{abstract}

\maketitle

\emph{Introduction}.---Quantum ergodicity is a fundamental concept explaining how unitary quantum evolution can lead to an equilibrium state described by statistical mechanics. While the eigenstate thermalization hypothesis (ETH)~\cite{Deutsch91,Srednicki94,D'Alessio16,Deutsch18} posits that \emph{generic} closed quantum many-body systems exhibit ergodicity, there are important exceptions to this paradigm, including many-body localized systems~\cite{Abanin19}, integrable systems~\cite{Calabrese16} (which are non-generic), dipole-conserving theories~\cite{Pai19,Khemani19,Sala19}, and a relatively new class of weakly nonergodic systems exhibiting ``quantum many-body scars" (QMBS)~\cite{Moudgalya18a,Turner17,Turner18,Moudgalya18b,Choi18,Khemani18,Lin18,Ho19,Ok19,Iadecola19}. 

Ergodicity breaking in such systems can often be attributed to the presence of symmetries (hidden, emergent or explicit) that preclude the establishment of a global equilibrium state. A notable exception arises in systems with QMBS, 
which exhibit nonergodic dynamics in the form of coherent oscillations of local observables after a quantum quench from certain initial states, as observed in a recent experiment in a Rydberg-atom quantum simulator~\cite{Bernien17}. In this case, the observed nonergodicity stems from the existence of an extensive set of special ``scarred" eigenstates that are unrelated to any symmetry of the Hamiltonian~\cite{Turner17}. This is a remarkable departure from the ETH scenario, wherein the finite-energy-density initial state would rapidly thermalize and lose coherence. The violation of ergodicity via scarring therefore presents a fundamental puzzle in our understanding of highly excited states in thermalizing systems that has spurred substantial recent interest.


The ubiquity and stability of QMBS are under active investigation. Multiple possible explanations of the underlying mechanism have been debated for the so-called ``PXP model" realized in the Rydberg experiment~\cite{Turner17,Turner18,Choi18,Khemani18,Lin18,Ho19,Shiraishi19,Iadecola19,Michailidis19,Moudgalya19}, ranging from proximity to integrability~\cite{Khemani18}, ``embedded" SU(2) dynamics~\cite{Choi18,Shiraishi17} and magnon condensation~\cite{Iadecola19}; moreover, connections have been made to gauge theory~\cite{Surace19}, symmetry-protected topological phases~\cite{Lin18,Shiraishi19}, and quantum Hall physics~\cite{Moudgalya19}. Given these various perspectives, it is highly desirable to find a tractable realization of scarring that can be established rigorously and its nonergodic properties studied analytically. While exact scarred eigenstates of the Affleck-Kennedy-Lieb-Tasaki (AKLT) spin chains~\cite{Affleck87} have been constructed analytically~\cite{Moudgalya18a,Moudgalya18b}, it is unclear whether and how these states lead to dynamical signatures resembling the experimental observations~\cite{Bernien17}.

In this paper, we study the spin-1 XY model on a hypercubic lattice in $d$ dimensions. We show that this well-known model surprisingly harbors an extensive set of anomalous scarred eigenstates at finite energy density that exhibit subextensive entanglement entropy and long-range space-time crystalline order~\cite{Wilczek12,Watanabe15,Sacha17}. These scarred states survive certain continuous deformations of the model and are eigenstates of an emergent SU(2) algebra that is \emph{not} part of the Hamiltonian's symmetry group. We further show that the scarred states lead to persistent oscillations of local observables following suitable quantum quenches. In particular, we show that quantum evolution starting from a suitable initial product state, prepared by applying a large symmetry-breaking field, shows perfect periodic revivals, while generic initial states rapidly thermalize.  Our results thus firmly establish the existence of QMBS in the spin-1 XY model.

\emph{Model}.---We study the spin-1 XY model
\begin{equation}\label{eq:H-XY}
H=J\sum_{\langle ij\rangle} \left(S^x_i S^x_j+S^y_i S^y_j\right)+h\sum_i S^z_i+D\sum_i \left(S^{z}_i\right)^2,
\end{equation}
where $S_i^\alpha$ $(\alpha=x,y,z)$ are spin-1 operators residing on the sites $i$ of a $d$ dimensional hypercubic lattice with volume $V=L^d$. We hereafter set $J=1$ and assume either periodic or open boundary conditions (PBC or OBC) as noted.  In~\cite{SM} we present a class of spin-$S$ generalizations of Eq.~\eqref{eq:H-XY} that also exhibit QMBS.

$H$ possesses a global U(1) symmetry generated by spin rotations about the $z$-axis and, depending on boundary conditions, may have translation and/or point-group symmetries.  For $d=1$ and OBC, $H$ also has a nonlocal SU(2) symmetry~\cite{Kitazawa03}, although this symmetry is \emph{not} requisite for scarring and can be removed, e.g., by adding $H_3=J_3\sum_{i} \left(S^x_i S^x_{i+3}+S^y_i S^y_{i+3}\right)$. In fact, any U(1)-symmetric exchange term preserving the bipartite structure of the hypercubic lattice is allowed~\cite{SM}. 

\begin{figure}[t!]
\begin{center}
\includegraphics[width=.85\columnwidth]{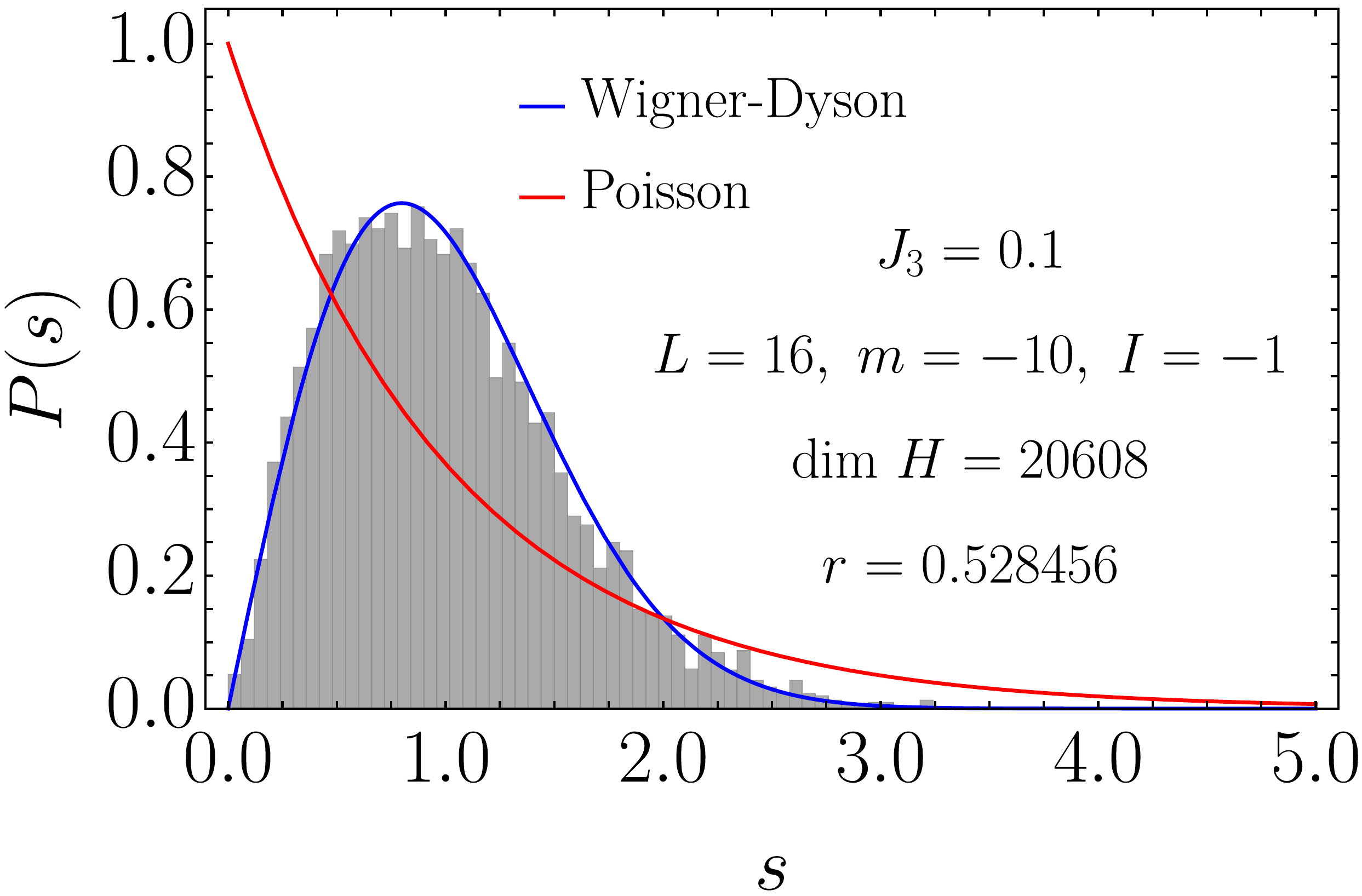}
\caption{Distribution of many-body level spacings $s$ in the middle half of the spectrum of $H$ for $d=1$ with open boundary conditions and $J_3=0.1$. The data are taken in the U(1) sector $\sum_i S^z_i=m=-10$ and the inversion sector $I=-1$. The $r$ value~\cite{Pal10} of the distribution given in the figure is close to the Wigner-Dyson result, $r_{\rm WD}\approx 0.5295$.}
\label{fig:1}
\end{center}
\end{figure}

The Hamiltonian $H$ is nonintegrable and, as we show in Fig.~\ref{fig:1}, the statistics of its many-body energy level spacings $s$ in a symmetry sector with sufficiently many levels follows the Wigner-Dyson (WD) distribution. WD level statistics is a common proxy for chaotic and ergodic behavior in quantum systems and indicates the absence of hidden or emergent symmetries that would strongly influence the level statistics (e.g., integrable systems follow the Poisson distribution shown for comparison in Fig.~\ref{fig:1}). 

It is well-known that WD level statistics alone is not sufficient to guarantee that the \emph{strong} ETH, positing that \emph{all} states in an energy window obey the ETH~\cite{Deutsch18}, holds. In special cases, a weak form of the ETH may hold~\cite{Biroli10} that allows for a rare set of anomalous eigenstates that violate the ETH. This possibility is remarkable in light of the fact that there is no protecting symmetry that prevents the anomalous states from mixing with thermal states at the same energy. Nevertheless, we now demonstrate that this scenario holds for the spin-1 XY model~\eqref{eq:H-XY}. The strong ETH fails due to the presence of the following athermal eigenstates:
\begin{equation}\label{eq:scar}
|\mathcal S_n\rangle= \mathcal{N}(n) \left(J^{+}\right)^n |\Omega\rangle,
\end{equation}
where $n=0,\dots,V$ is an integer, $|\Omega\rangle=\bigotimes_i |m_i=-1\rangle$ is the fully-polarized ``down" state, $m_i=-1,0,1$ are the eigenvalues of $S^z_i$, $\mathcal{N}(n)=\sqrt{\frac{(V-n)!}{n!V!}}$ are normalization factors, and
\begin{equation}\label{eq:Jpm}
J^{\pm}= \frac{1}{2}\sum_i e^{i \bm{r}_i\cdot \bm{\pi}} \left(S^{\pm }_i\right)^2.
\end{equation}
In Eq.~\eqref{eq:Jpm}, $\bm{r}_i$ are the spins' coordinates and $\bm{\pi}=(\pi,\pi,\dots,\pi)$.  The state $|\mathcal S_n\rangle$ contains $n$ \emph{bimagnons} (i.e.,~doubly-raised spins), each with momentum $\bm k=\bm\pi$.  In~\cite{SM}, we show that they are frustration-free eigenstates of the Hamiltonian~\eqref{eq:H-XY} with energy $E_n=h(2n-V)+VD$ and total magnetization $m_n=2n-V$. There we also highlight another (orthogonal) tower of exact eigenstates that arise for PBC and $D=0$ in $d=1$.


Interestingly, the operators $J^\pm$ are generators of an SU(2) algebra (distinct from that of Ref.~\cite{Kitazawa03}) defined by
\begin{equation}\label{eq:su(2)}
J^z=\frac{1}{2}\sum_i S^z_i;\,\,\,\left[J^+,J^-\right]=2J^z;\,\,\,\left[J^z,J^\pm\right]=\pm J^\pm.
\end{equation}
Note that the spin-1 nature of the microscopic spins is crucial for this algebra to hold. These $SU(2)$ generators do not all commute with $H$: $\left[H,J^\pm\right]\neq 0$ while $[H,J^z]=0$. Nevertheless, the scarred states~\eqref{eq:scar}
form a representation of this emergent SU(2) algebra with spin $j=V/2$ (the maximum possible value): 
\begin{equation}\label{eq:spin}
\bm J\cdot \bm J\, |\mathcal S_n\rangle=\frac{V}{2}\left(\frac{V}{2}+1\right)\mathcal |\mathcal S_n\rangle,
\end{equation}
where $\bm J\cdot \bm J=\frac{1}{2}\left(J^+J^- + J^-J^+\right)+\left(J^z\right)^2$. $J^{\pm}$ thus act as ladder operators for the scarred states:
\begin{equation}\label{eq:ladder}
J^{\pm}|\mathcal S_n\rangle=\sqrt{j(j+1)-\frac{m_n}{2}\left(\frac{m_n}{2}\pm1\right)}|\mathcal S_{n\pm1}\rangle,
\end{equation}
where $j=V/2$ and $m_n/2=n-V/2$.

\begin{figure}[t!]
\begin{center}
\includegraphics[width=.95\columnwidth]{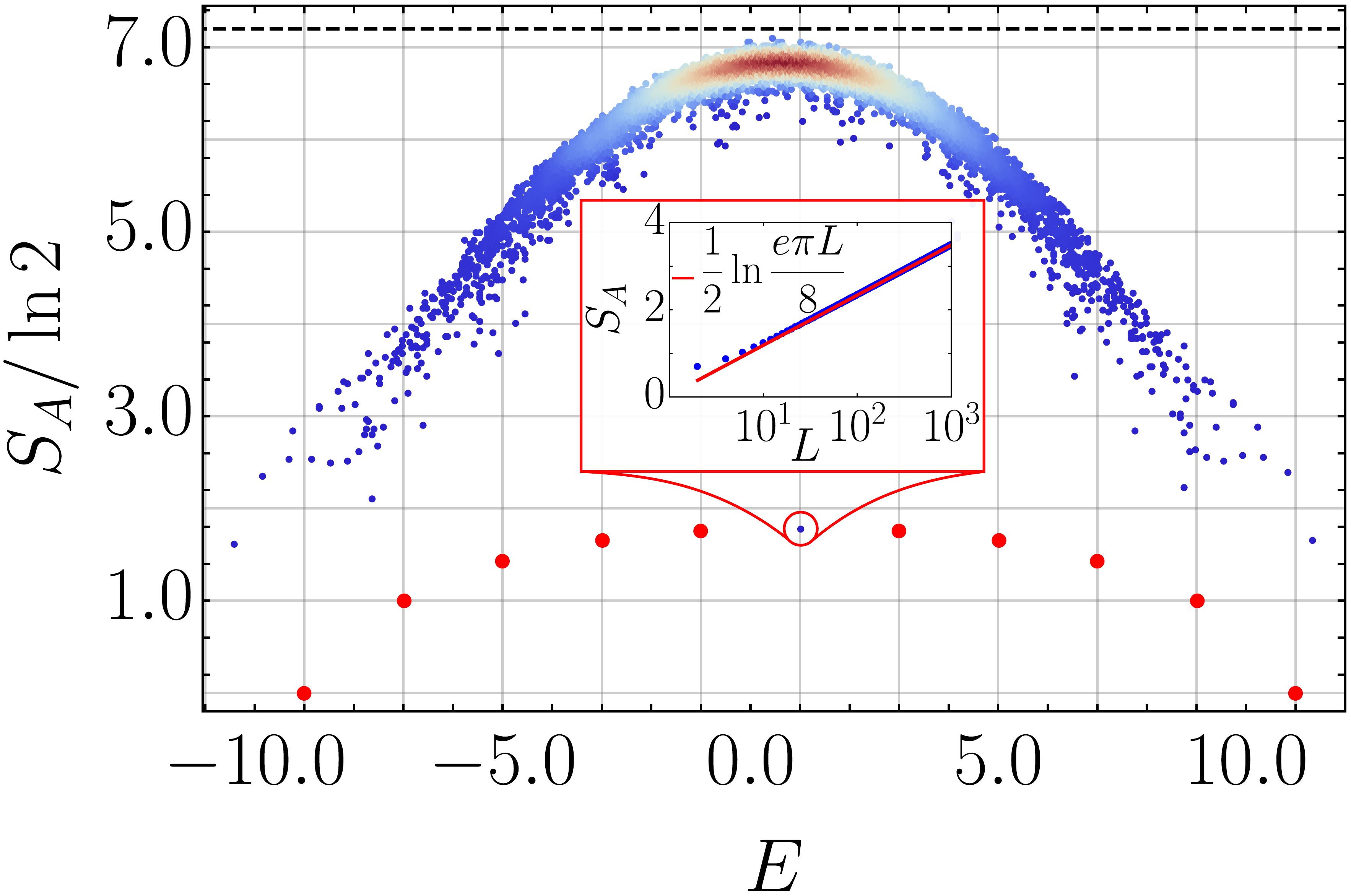}
\caption{Bipartite entanglement entropy $S_A$ of eigenstates of $H$ for $d=1$, $L=10$ and $(h,D,J_3)=(1,0.1,0.1)$ with OBC. States in the zero-magnetization sector (smaller points) are color coded by the density of states (warmer colors imply higher density). The dashed line at $S_A^{\rm ran}=\frac{L}{2}\ln 3-\frac{1}{2}$ indicates $S_A$ for a random state. Larger red points indicate scarred states~\eqref{eq:scar} in U(1) sectors with $m_n\neq0$. Inset: $S_A$ for $|\mathcal S_{L/2}\rangle$ as a function of $L$, cf. Eq.~\eqref{eq:log}.}
\label{fig:2}
\end{center}
\end{figure}

The scarred states~\eqref{eq:scar} at generic $n$ are not thermal even though they have finite energy density and reside in symmetry sectors with exponentially many states; hence they violate ETH.  To show this, we first consider their bipartite entanglement entropy  $S_A=-{\rm tr}\rho_A\ln\rho_A$ where $\rho_A$ is the reduced density matrix for a region $A$ of size $V_A=V/2$. We plot $S_A$ vs.~energy $E$ for eigenstates in the zero-magnetization sector in Fig.~\ref{fig:2}, highlighting the lightly entangled scarred state $|S_{V/2}\rangle$ with a red circle (eigenstates are obtained using exact diagonalization for $d=1$, $L=10$). ETH-obeying states have extensive ``volume-law" entanglement entropy, $S_A\propto V$.
For states near the middle of the spectrum (nominally at infinite temperature), $S_A$ should approach the value for a random state, $S_A^{\rm ran}=\frac{V}{2}\ln3-\frac{1}{2}$~\cite{Page93} (dashed line in Fig.~\ref{fig:2}), which appears to be approximately true for a large fraction of states near $E\approx0$.
Thus, to show that the states~\eqref{eq:scar} violate the ETH, we need only show that their entanglement entropy is subextensive.

The simplicity of the states~\eqref{eq:scar} allows for an analytical calculation of the full entanglement spectrum from which the entanglement entropy can be obtained, see~\cite{SM}.  The resulting values of $S_A$ for the state $|\mathcal S_{V/2}\rangle$ are plotted in the inset of Fig.~\ref{fig:2} as a function of system length $L$ for $d=1$. In fact, one can show analytically that $S_A$ takes the asymptotic form~\cite{SM}
\begin{equation}\label{eq:log}
S_A(n=V/2)
\ \underset{V\to\infty}{\xrightarrow{\hspace*{.75cm}}}\
\frac{1}{2}\left(\ln \frac{\pi V}{8}+1\right),
\end{equation}
cf.~inset to Fig.~\ref{fig:2}. The state $|\mathcal S_{V/2}\rangle$ has the highest entanglement of all scarred states~\eqref{eq:scar} (cf.~Fig.~\ref{fig:2}), so Eq.~\eqref{eq:log} demonstrates conclusively that these states exhibit subextensive entanglement entropy scaling at most logarithmically with system size.



It is instructive to compare the scarred states~\eqref{eq:scar} with other examples of exact excited states of nonintegrable models, in particular the ``$\eta$-pairing" states of the Hubbard model~\cite{Yang89} and the scarred states of the AKLT chain~\cite{Moudgalya18a}. Both of the latter examples also host towers of states with logarithmic entanglement~\cite{Vafek17,Moudgalya18b} obtained by acting repeatedly with some operator on a parent state. The $\eta$-pairing example is unique in that it is protected by ``$\eta$ symmetry," i.e.,~the analogues of $J^\pm$ are eigenoperators of the Hamiltonian and the $\eta$-pairing states are the only states in their respective symmetry sectors. Thus, the $\eta$-pairing states are neither ETH-violating nor bona fide scarred states (despite many similar features). The AKLT scarred states \emph{do} violate the ETH, and interestingly, are created by the same operator, $J^+$, as in Eqs.~(\ref{eq:scar}--\ref{eq:Jpm}). However, the parent state in that case is the AKLT ground state rather than the fully polarized state $|\Omega\rangle$. This is crucial because the AKLT scarred states \emph{do not} form a representation of the SU(2) algebra~\eqref{eq:su(2)}. It is an important outstanding question whether such a structure exists for the AKLT model, as it could be used to determine the dynamical signatures of the scarred states, which (to the best of our knowledge) remain unknown. For the scarred states presented here this is not the case, and we now show that their dynamical signatures can be deduced directly from the SU(2) algebra~\eqref{eq:su(2)}.

\emph{Space-Time Crystalline Order}.---We first demonstrate the presence of off-diagonal long-range order (ODLRO)~\cite{Yang62} in the scarred states associated with the condensation of bimagnons at momentum $\bm{\pi}$.  Such order is also present in the $\eta$-pairing states, where it is indicative of superconductivity~\cite{Yang89}. Here, the order is of a \emph{spin-nematic} nature: the order parameter $O_{\bm{q}}=\frac{1}{V}\sum_{i}e^{i\bm{r}_i\cdot \bm{q}}\left(S^+_i\right)^2$ has long-range connected correlations at wavevector $\bm{q}=\bm{\pi}$ in the scarred states. This is indicated by a finite value of the correlation function $\langle \mathcal S_n|O^\dagger_{\bm\pi}O_{\bm\pi} |\mathcal S_n\rangle$ [note $\langle \mathcal S_n|O^\dagger_{\bm\pi}|\mathcal S_n\rangle=0$ by U(1) symmetry].
Using Eqs.~(\ref{eq:Jpm}), (\ref{eq:ladder}) one immediately obtains
\begin{equation}\label{eq:ODLRO}
\langle \mathcal S_n|O^\dagger_{\bm\pi}O_{\bm\pi} |\mathcal S_n\rangle= 1-m^{\prime 2}_n + \mathcal O(1/V),
\end{equation}
where the $\mathcal O (1/V)$ terms vanish in the limit $V\to\infty$ and $m_n{^\prime}=m_n/V$ is the magnetization density. We thus find that the scarred states  $|\mathcal S_n\rangle$ (aside from the zero-measure set with $m^\prime=\pm 1$) possess spin-nematic ODLRO. This implies that the spin \emph{fluctuations} in the $x$-$y$ plane break the U(1) spin-rotation symmetry spontaneously without long-range magnetic order (i.e., time-reversal symmetry is preserved).  This remarkable property also heralds ETH violation: ODLRO is impossible for ETH-obeying states in the middle of the spectrum [such states are nominally at infinite temperature, where the thermal density matrix $\rho=e^{-\beta H}$ in a given U(1) sector is trivial]. 

The ODLRO in Eq.~\eqref{eq:ODLRO} immediately implies that the 
scarred states also support long-range \emph{spacetime} correlations, the
defining characteristic of space-time crystalline order~\cite{Watanabe15,Khemani17}.
Up to $1/V$ corrections we have
\begin{equation}\label{eq:STCO}
{\rm Re}\, \langle \mathcal S_n|O^\dagger_{\bm\pi}(t)O_{\bm\pi}(0) |\mathcal S_n\rangle= (1-m^{\prime 2}_n)\cos (2ht).
\end{equation}
This space-time crystalline order can ultimately be traced back to the condensation of $\pi-$bimagnons. We note that the existence of this order does not violate the no-go theorems establishing its impossibility at thermal equilibrium~\cite{Bruno13,Watanabe15}; since the scarred states violate the ETH, these no-go theorems do not apply.

\begin{figure}[t!]
\begin{center}
\includegraphics[width=.9\columnwidth]{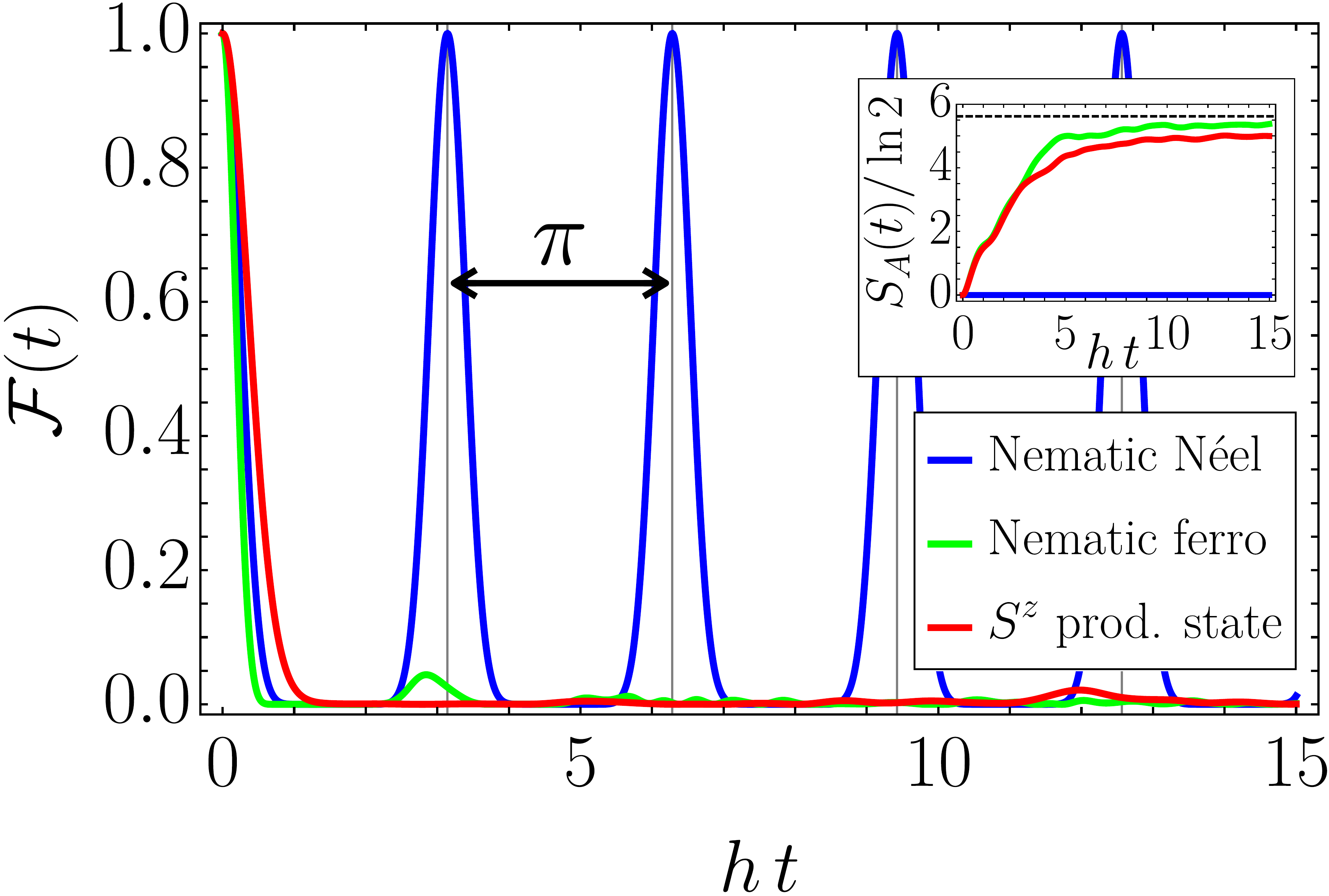}
\caption{Many-body fidelity $\mathcal F(t)=\left|\langle \psi(0)|\psi(t)\rangle\right|^2$ for various initial states ($d=1$, $L=8$, and remaining parameters as in Fig.~\ref{fig:2}). The nematic N\'eel initial state, Eq.~\eqref{eq:psi0}, exhibits perfect revivals described by Eq.~\eqref{eq:fidelity}, while generic initial states decay rapidly. Inset: Entanglement dynamics after a quench, showing that generic initial states lead to rapid entanglement growth and saturation near the value for a random state (dashed line), while the special initial state does not.}
\label{fig:3}
\end{center}
\end{figure}

\begin{figure}[t!]
\begin{center}
\includegraphics[width=0.5\columnwidth]{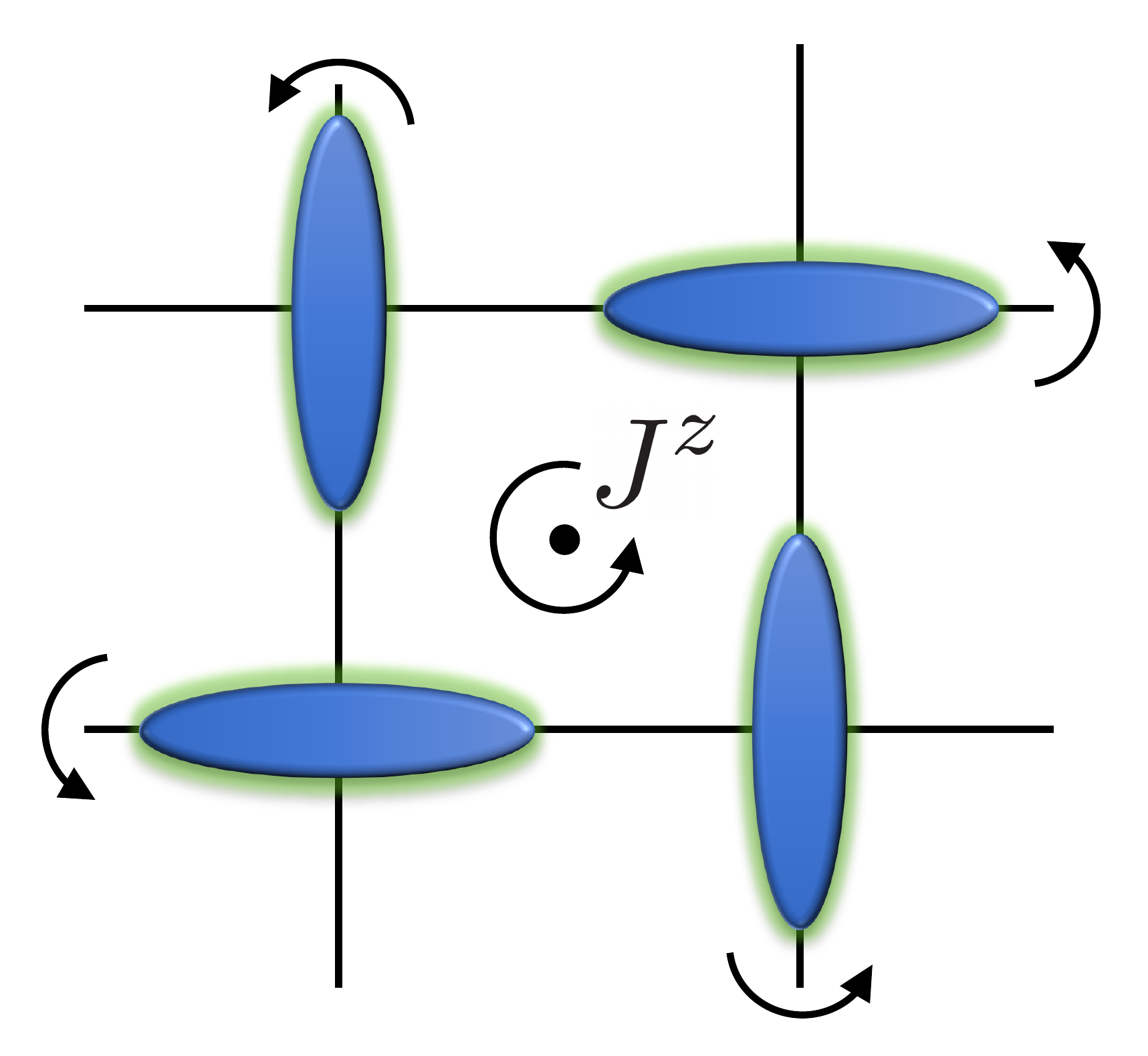}
\caption{Schematic of the spin-nematic order parameter on a 2d square lattice precessing around the applied field, $\sum_i S^z_i\propto J^z$. The staggered local directors (blue cigars) are synchronized such that their dynamics stay within the manifold of scarred states, cf. Eqs.~(\ref{eq:init}-\ref{eq:director}). }
\label{fig:4}
\end{center}
\end{figure}

\emph{Dynamical Signature of Scars}.---We now demonstrate that the eigenstate properties of $|\mathcal S_n\rangle$ derived above have significant consequences for the dynamics of local observables after certain quantum quenches. To illustrate, we initialize the system in the ground state $|\psi_0\rangle$ of the staggered rhombic anisotropy Hamiltonian
\begin{equation}\label{eq:quench}
H_A=\frac{1}{2}\sum_i e^{i\bm{r}_i\cdot\bm{\pi}}\left(\left(S^{x}_i\right)^2-\left(S^{y}_i\right)^2\right).
\end{equation}
This Hamiltonian is relevant to scarring since it can be rewritten in the form $H_A=\frac{1}{2}\left(J^++J^-\right)\equiv J^x$. $|\psi_0\rangle$ is thus the lowest-weight state of $J_x$ in the spin-$V/2$ representation of the SU(2) algebra~\eqref{eq:su(2)}, which we call the ``nematic N\'eel" state
\begin{equation}\label{eq:psi0}
|\psi_0\rangle = \bigotimes_i\left(\frac{|m_i=+1\rangle-e^{i\bm{r}_i\cdot\bm{\pi}}|m_i=-1\rangle}{\sqrt{2}}\right).
\end{equation}
Since this is an eigenstate of the spin-$V/2$ representation of Eq.~\eqref{eq:su(2)}, it resides entirely within the scarred manifold,
\begin{equation}\label{eq:init}
|\psi_0\rangle=\sum_{n=0}^{V}c_n |\mathcal S_n\rangle,\,\,\,c_n^2=\frac{1}{2^{V}}\left(\begin{array}{c}V\\n\end{array}\right).
\end{equation}
The fidelity of this initial state under evolution with $H$ is thus given by
\begin{equation}\label{eq:fidelity}
\mathcal F_0(t)=\left|\langle\psi_0|\psi_0(t)\rangle\right|^2=\cos^{2 V}(ht),
\end{equation}
which exhibits perfect revivals with period $T=\pi/h$. As a result, all local observables oscillate with the revival period.  This behavior is shown in Fig.~\ref{fig:3} where we compare fidelities of other initial states that decay rapidly. The inset to Fig.~\ref{fig:3} shows entanglement dynamics after a quench: while generic product states rapidly approach maximal entanglement, the special initial state $|\psi_0\rangle$ remains a product state under time evolution with $H$. Indeed, this evolution merely imparts phase factors $e^{\mp iht}$ to the terms $|m_i=\pm1\rangle$ in Eq.~\eqref{eq:psi0}. Physically this corresponds to a set of spin-nematic \emph{directors} precessing in the $x$-$y$ plane with frequency twice the applied field, see Fig.~\ref{fig:4}. The local director angle $\theta_i$ may be defined in terms of the phase of the local order parameter
\begin{equation}\label{eq:director}
O_i=\left(S^+_i\right)^2.
\end{equation}
Time evolution yields $\langle \psi_0(t)|O_i|\psi_0(t)\rangle=e^{2iht}\equiv|O_i|e^{2i\theta_i}$ and hence the phase winds as $\theta_i=ht$ with $|O_i|=1$.  The directors thus oscillate coherently and in a synchronized fashion when initially staggered in space, as in Eq.~\eqref{eq:psi0} and shown schematically in Fig.~\ref{fig:4}. 

Crucially, this dynamical behavior \emph{does not} originate from a set of freely precessing directors. If it did, the ``nematic ferro'' state (with directors aligned) would also show oscillations, and this is clearly not the case (cf.~Fig.~\ref{fig:3}). Rather, the observed revivals originate from the precession of a single emergent \emph{macroscopic} staggered director.  The existence of this director is enabled by the long-range connected correlations in the scarred states $|\mathcal S_n\rangle$, which in turn originate from the emergent SU(2) algebra~\eqref{eq:su(2)}.

Finally, we note that the scarred states persist in the presence of the staggered rhombic anisotropy $H_A$, Eq.~\eqref{eq:quench}. Because $H_A\propto J^x$, it cants the effective magnetic field about which the macroscopic director precesses (indeed, more generally one can add an additional anisotropy $H_B\propto J^y$). The resulting scarred eigenstates can be obtained from the states $|\mathcal S_n\rangle$ by an appropriate SU(2) rotation using the generators~\eqref{eq:su(2)}. This observation implies that one need not quench the staggered anisotropy in order to observe persistent oscillations. In a system with \emph{fixed} staggered rhombic anisotropy as in Eq.~\eqref{eq:quench}, one may instead polarize the initial state by applying a large homogeneous magnetic field $\propto J^z$. The dynamics of the fully polarized state will then show persistent oscillations due to the presence of the rotated scarred states, whereas generic states will not.

 
\emph{Conclusion}.---In this paper we uncovered a set of exact scarred eigenstates in nonintegrable spin-1 XY magnets, leading to weak ergodicity breaking and strong-ETH violation. These states have properties that are impossible in ETH-obeying states, including subextensive entanglement entropy and spin-nematic ODLRO. The scarred states are maximal-spin eigenstates of an emergent SU(2) algebra that does not commute with the Hamiltonian. Using this algebra we showed that the scarred states enable coherent many-body revivals following suitable quantum quenches.

This work provides a novel context in which several hypothesized characteristics of QMBS in the Rydberg-atom quantum simulator~\cite{Bernien17} become exact.  For example, in Ref.~\cite{Choi18} it was suggested that an emergent SU(2) algebra could be responsible for the observed revivals, while Ref.~\cite{Iadecola19} numerically demonstrated ODLRO and spacetime crystalline order in the scarred states. Thus the exact scarred states uncovered here suggest a common paradigm for QMBS that could be relevant across various physical models and which can be compared with other exact mechanisms leading to strong-ETH violation, including embedded Hilbert spaces~\cite{Shiraishi17,Ok19,Shiraishi19,SM} and emergent invariant subspaces~\cite{Iadecola18,Khemani19,Sala19}.  This work also opens the possibility of searching experimentally for scarred dynamics in physical XY magnets with appropriate single-ion anisotropies, or engineering it in superconducting circuits that could simulate spin-1 XY models.

\acknowledgements{We thank C.-J.~Lin for pointing out the connection between our results and those of Ref.~\cite{Shiraishi17}, on which we comment further in~\cite{SM}. This work is supported by Microsoft and the Laboratory for Physical Sciences. T.I. acknowledges a JQI postdoctoral fellowship.}

\bibliography{refs_zero}

\begin{widetext}

\begin{center}
\textbf{\large Supplemental Material for ``Weak Ergodicity Breaking and Quantum Many-Body Scars in Spin-1 XY Magnets"}
\end{center}

\section{Appendix A: Derivation of Exact Athermal Eigenstates}
\label{sec: Appendix A}

\setcounter{equation}{0}
\renewcommand{\theequation}{A\arabic{equation}}

In this Appendix we show that the scar states $|\mathcal S_n\rangle$ defined in Eq.~\eqref{eq:scar} are indeed eigenstates of the Hamiltonian $H$ of Eq.~\eqref{eq:H-XY} on a $d$-dimensional hypercubic lattice. We also show that these states remain in the presence of any exchange term that preserves the original U(1) symmetry of Eq.~\eqref{eq:H-XY} and the bipartiteness of the hypercubic lattice.  Finally, we comment on the existence of other exact eigenstates in the many-body spectrum of $H$.  These additional exact eigenstates include another tower of athermal eigenstates at finite energy density that arise only for periodic boundary conditions and anisotropy $D=0$.

\subsection{The Scar States $|\mathcal S_n\rangle$}

To demonstrate that the states $|\mathcal S_n\rangle$ are indeed eigenstates of $H$, we work in the local $S^z$ basis, where we label the local spin states
\begin{align}
|m_i=\pm 1\rangle,|m_i=0\rangle \equiv |\pm\rangle,|0\rangle,
\end{align}
respectively.  The states are then automatically eigenstates of the $h$ and $D$ terms, and it remains to check the action of the nearest-neighbor hopping term $J$ for each $n$. We do this below for the case $n=1$ in Sec.~\ref{sec:NN-exch} before proceeding to consider further-neighbor exchange processes in Sec.~\ref{sec:FN-exch}.  We then consider the case $n>1$ in Sec.~\ref{sec:hardcore} and show that the presence of additional bimagnons does not spoil the hopping properties demonstrated in the preceding subsections, a fact that we attribute to the hardcore nature of the bimagnons. 

\subsubsection{$|\mathcal S_1\rangle$ Nullifies Nearest-Neighbor Exchange}
\label{sec:NN-exch}

We begin by considering $d=1$, where the state
\begin{align}
\label{appeq:S_1d=1}
|\mathcal S_1\rangle=\frac{1}{\sqrt{L}}\sum^L_{j=1} e^{i \pi j}\, |\,\underset{j-1}{\underbrace{-\cdots}}+\underset{L-j}{\underbrace{-\cdots}}\,\rangle
\end{align}
(recall $V=L$ in $d=1$).
We will assume periodic boundary conditions (PBC), although we will see later that this is not necessary. We now consider the action of 
\begin{align}
\label{appeq:H1}
H_1=\sum^L_{i=1}\(S^x_iS^x_{i+1}+S^y_iS^y_{i+1}\)
\end{align}
on the state~\eqref{appeq:S_1d=1}, namely
\begin{align}
H_1|\mathcal S_1\rangle\propto \sum^L_{j=1} e^{i \pi j}\bigg(|\,\underset{j-1}{\underbrace{-\cdots}}\,00\underset{L-j-1}{\underbrace{-\cdots}}\,\rangle+|\,\underset{j-2}{\underbrace{-\cdots}}\,00\,\underset{L-j}{\underbrace{-\cdots}}\,\rangle\bigg).
\end{align}
We now take advantage of PBC to shift $j\to j+1$ in the sum for the second term in parentheses above, giving
\begin{align}
\label{appeq:H1S1=0d=1}
H_1|\mathcal S_1\rangle\propto \(1+e^{i\pi}\)\sum^L_{j=1} e^{i \pi j}\, |\,\underset{j-2}{\underbrace{-\cdots}}\,00\,\underset{L-j}{\underbrace{-\cdots}}\,\rangle=0.
\end{align}
Evidently, $|\mathcal S_1\rangle$ is a zero-energy eigenstate of $H_1$. This result relies on the fact that the bimagnon in the state $\mathcal S_1$ has momentum $\pi$.  Indeed, replacing $\pi$ in Eq.~\eqref{appeq:S_1d=1} by $p(2\pi/L)$ for arbitrary $p\in\mathbbm Z$ does not yield an eigenstate, even at the single-bimagnon level.  This indicates that we should view the $\pi$-bimagnon as a stable bound state of two magnons, which we define as isolated ``$0$"s in a background of ``$-$"s.

Eqs.~\eqref{appeq:S_1d=1}--\eqref{appeq:H1S1=0d=1} generalize readily for $d>1$, where
\begin{align}
|\mathcal S_1\rangle=\frac{1}{\sqrt{V}}\sum_{i} e^{i \bm{r}_i\cdot \bm{\pi}}\,
\bigotimes_j
\begin{cases} 
\left|+\right\rangle & j=i\\
\left|-\right\rangle & \text{otherwise}
\end{cases}
\end{align}
and
\begin{align}
\label{appeq:H1-decomp}
H_1=\sum_{\langle i,j \rangle}\(S^x_iS^x_j+S^y_iS^y_j\)=\sum^d_{a=1}H^a_1,
\end{align}
where $a=1,\dots,d$ denote the principal axes of the hypercube and $H^a_1$ contains nearest-neighbor exchange terms acting only along the principal axis $a$. It then suffices to consider the action of $H^a_1$ on $|\mathcal S_1\rangle$ for each $a=1,\dots,d$.  These calculations proceed identically to the $d=1$ case, with the only modification being that the prefactor in the analog of Eq.~\eqref{appeq:H1S1=0d=1} is replaced by $1+e^{i\hat{\bm{e}}_a\cdot\bm \pi}=0$, where $\hat{\bm{e}}_a$ is the unit vector in the $a$ direction.

\subsubsection{$|\mathcal S_1\rangle$ Nullifies Certain Further-Neighbor Exchanges}
\label{sec:FN-exch}

We next consider a family of further-neighbor exchange terms,
\begin{align}
\label{appeq:Hm}
H_{\bm m}=\sum_{i,j}\delta_{j,N_i(\bm m)}\(S^x_iS^x_j+S^y_iS^y_j\),
\end{align}
where $\bm m = (m_a)^{d}_{a=1}$ is a $d$-dimensional vector with integer entries and $N_i(\bm m)$ is the site obtained by translating $\bm r_i\to\bm r_i+\bm m$. The Hamiltonian $H_{\bm m}$ can be divided into a set of terms acting along parallel lines, where each lattice site $i$ belongs to exactly one line defined as the one containing the sites $i$ and $N_i(\bm m)$.  Along each such line, we again have a Hamiltonian of the form \eqref{appeq:H1} acting along an effectively one-dimensional sublattice of the hypercube.  Using this decomposition of $H_{\bm m}$ and an analogous decomposition of $|\mathcal S_1\rangle$, one can compute the analog of Eq.~\eqref{appeq:H1S1=0d=1} and in particular the analogous prefactor.  One finds a prefactor
\begin{align}
\label{appeq:m-crit}
1+e^{i\bm m\cdot\bm{\pi}}=1+e^{i\, \pi\sum^d_{a=1}m_a}=0 \iff \sum^d_{a=1} m_a = 1\ \text{mod}\ 2.
\end{align}
We arrive at the conclusion that any exchange term $H_{\bm m}$ of the form \eqref{appeq:Hm} on a $d$-dimensional hypercube for which $\bm m$ satisfies Eq.~\eqref{appeq:m-crit} annihilates the state $|\mathcal S_1\rangle$.  An example of such an exchange term term in $d=1$ is the third-neighbor hopping term used to obtain Fig.~\ref{fig:1}.

The constraint on $\bm m$ in Eq.~\eqref{appeq:m-crit} amounts to the requirement that $H_{\bm m}$ preserves the bipartiteness of the hypercubic lattice.  In other words, if we label the two sublattices of the hypercubic lattice by $\mathcal A$ and $\mathcal B$, Eq.~\eqref{appeq:m-crit} implies that $H_{\bm m}$ connects only $\mathcal A$ sites to $\mathcal B$ sites, and never two sites on the same sublattice.  This is because Eq.~\eqref{appeq:m-crit} essentially states that any path in the lattice connecting any two points separated by the vector $\bm m$ must traverse an odd number of links.  

Combining this observation with the fact that the bimagnons are hardcore objects (see Sec.~\ref{sec:hardcore}), we expect (but have not proven) that it is possible to generalize the construction of this paper to arbitrary bipartite graphs by replacing the factor $e^{i\bm r_i\cdot\bm\pi}$ in Eq.~\eqref{eq:Jpm} with a factor $s_i=\pm 1$, where the choice of sign is based on whether the site $i$ belongs to subgraph $\mathcal A$ or $\mathcal B$.

\subsubsection{$n>1 \!:$ The Importance of Being Frustration-Free}
\label{sec:hardcore}

Having demonstrated that the state $|\mathcal S_1\rangle$ is a zero-energy eigenstate of a family of exchange Hamiltonians [including the nearest-neighbor one appearing in Eq.~\eqref{eq:H-XY}], it remains to show that the same is true for the states $|\mathcal S_n\rangle$ for $n>1$.  It suffices to show that this is the case for $d=1$ since, as we have explained in Secs.~\ref{sec:NN-exch} and \ref{sec:FN-exch}, the family of exchange Hamiltonians we consider can be viewed as acting along independent one-dimensional subsystems for general $d$.

For $d=1$, we have
\begin{align}
\label{appeq:Sn}
|\mathcal S_n\rangle=\binom{L}{n}^{-\frac{1}{2}}\sum_{i_1\neq\dots\neq i_n}(-1)^{i_1+\dots+i_n}\, \bigotimes^{L}_{j=1}
\begin{cases} 
\left|+\right\rangle & j\in\{i_1,\dots,i_n\}\\
\left|-\right\rangle & \text{otherwise}
\end{cases}.
\end{align}
Importantly, $|\mathcal S_n\rangle$ contains all possible permutations of the $n$ bimagnons with weights given by an alternating sign.
We first consider the nearest-neighbor exchange Hamiltonian $H_1$, which we write as a sum of bond terms,
\begin{align}
\label{appeq: H1 local}
H_1=\sum_{i}h^1_{i,i+1}.
\end{align}
We will show that $|\mathcal S_n\rangle$ is a \textit{frustration-free} zero-energy eigenstate of $H_1$, i.e. that $h^1_{i,i+1}|\mathcal S_n\rangle=0$ for all $i$.

We proceed one bond at a time.  Any spin configuration in $|\mathcal S_n\rangle$ for which the spins on sites $i,i+1$ are in the $++$ or $--$ state is automatically annihilated by $h^1_{i,i+1}$.  This is a manifestation of the inertness of the vacuum state $|\Omega\rangle=|-\cdots-\rangle$ and the fact that the bimagnons are hardcore objects, i.e., two of them cannot reside on the same site.  This leaves the set of configurations for which the spins on sites $i,i+1$ are in the $+-$ and $-+$ states.  Because Eq.~\eqref{appeq:Sn} is an antisymmetrized sum over all configurations of $n$ $``+"s$ on a background of $``-"s$, any configuration $|\mathcal C\rangle=|L\rangle\otimes|+-\rangle\otimes|R\rangle$ has a corresponding configuration $|\tilde{\mathcal C}\rangle=|L\rangle\otimes|-+\rangle\otimes|R\rangle$ appearing in $|\mathcal S_n\rangle$ with opposite sign.
(Here, $|L\rangle$ and $|R\rangle$ are \textit{fixed} configurations of $``+"$s and $``-"$s.)  Since $h^1_{i,i+1}|+-\rangle=h^1_{i,i+1}|-+\rangle=|00\rangle$, $h^1_{i,i+1}$ maps both $|\mathcal C\rangle$ and $|\tilde{\mathcal C}\rangle$ to the same configuration, $|L\rangle\otimes|00\rangle\otimes|R\rangle$.  Since $|\mathcal C\rangle$ and $|\tilde{\mathcal C}\rangle$ appear in $|\mathcal S_n\rangle$ with opposite signs, $h^1_{i,i+1}$ annihilates the pair of configurations $|\mathcal C\rangle$ and $|\tilde{\mathcal C}\rangle$. It immediately follows that $H_1|\mathcal S_n\rangle=0$.

Three comments are in order. First, note that the argument above demonstrates that the eigenstates $|S_n\rangle$ are insensitive to whether periodic or open boundary conditions are imposed: the difference between the two cases is simply whether or not one includes the term $h^1_{L,1}$ in $H_1$, and this term would again annihilate $|\mathcal S_n\rangle$ by the same argument.  Second, note that the argument above generalizes in a straightforward manner to any exchange term connecting sites $i$ and $i+2p+1$, i.e., any odd-neighbor hopping.  In this case, one considers sites a distance $2p+1$ apart, and again the configurations with $``+"$ and $``-"$ exchanged come with opposite signs.  Third, note that the frustration-free nature of these eigenstates means that they remain exact eigenstates in the presence of arbitrary bond disorder in the exchange couplings.

\subsection{Other Notable Eigenstates}

We now point out the existence of additional exact eigenstates in the spectrum of $H$.  Some of these states we consider ``special," while others we do not.  The designation ``special" is somewhat arbitrary---for the present purposes we take it to mean that the athermal states in question contain a finite density of excitations so that the number of states in their corresponding symmetry sector grows exponentially with system size.  We reserve this designation for such states because the states with a vanishing density of excitations above the fully polarized states $|-\cdots-\rangle$ and $|+\cdots+\rangle$, which we discuss in Sec.~\ref{sec:non-special}, can be smoothly connected to low-lying excited states by applying, e.g., a sufficiently large magnetic field $h$, and thus have a small number of states in their symmetry sector. 

\subsubsection{Exact But Not ``Special" Eigenstates}
\label{sec:non-special}

The Hamiltonian \eqref{eq:H-XY} admits a series of single-magnon (i.e., single-``0") excitations atop the fully polarized states $|\Omega\rangle=|-\cdots-\rangle$ and $|\Omega'\rangle=|+\cdots+\rangle$.  These states and their energy eigenvalues are given by
\begin{subequations}
\begin{align}
|\Omega,\bm k\rangle&=\frac{1}{\sqrt{2V}}\sum_{i} e^{i\bm{r}_i\cdot \bm k}\, S^+_i|\Omega\rangle,
\indent
E_{\Omega,\bm k}=2\sum^d_{a=1}\cos k_a-h(V-1)+D(V-1)
\\
|\Omega',\bm k\rangle&=\frac{1}{\sqrt{2V}}\sum_{i} e^{i\bm{r}_i\cdot \bm k}\, S^-_i|\Omega'\rangle,
\indent
E_{\Omega',\bm k}=2\sum^d_{a=1}\cos k_a+h(V-1)+D(V-1),
\end{align}
\end{subequations}
where the momenta take the allowed values $k_a = \frac{2\pi}{L}\, p$ ($p=0,\dots,L-1$).  Their existence is guaranteed for arbitrary $U(1)$-symmetric exchange terms, as the action of such terms is simply to propagate the magnon across the lattice.

\subsubsection{Another Tower of ``Special" Eigenstates for $d=1$}
\label{sec:special}

Intriguingly, for $d=1$ there is another tower of generically finite-energy-density eigenstates consisting of effectively independent excitations with momentum $\pi$.  This tower of states is given (up to normalization) by
\begin{align}
\label{appeq:S'n}
|\mathcal S^\prime_n\rangle \propto \sum_{i_1\neq i_2\neq\dots\neq i_n}(-1)^{i_1+\dots+i_n}\, (S^+_{i_1}S^+_{i_1+1})(S^+_{i_2}S^+_{i_2+1})\dots(S^+_{i_n}S^+_{i_n+1})\, |\Omega\rangle.
\end{align}
The states $|\mathcal S^\prime_n\rangle$ contain $n$ ``bond" bimagnons, which consist of \textit{pairs} of spin-flips against a polarized background, i.e., $|-\cdots00-\cdots\rangle$.
Similarly to the tower of states discussed in the main text, there are $L+1$ such states, starting from $|\mathcal S^\prime_0\rangle=|\Omega\rangle$ and ending with $|\mathcal S^\prime_L\rangle=|\Omega'\rangle$.  However, apart from these lowest- and highest-weight states, the tower~\eqref{appeq:S'n} is orthogonal to the tower~\eqref{eq:scar}.
Moreover, the states in Eq.~\eqref{appeq:S'n} are less robust than those discussed in the main text: they exist only for PBC and $D=0$ in Eq.~\eqref{eq:H-XY}.  The state $|\mathcal S^\prime_n\rangle$ resides in the same total magnetization sector as $|\mathcal S_n\rangle$ and has energy
\begin{align}
E^\prime_n=h(2n-V).
\end{align}
Unlike the states $|\mathcal S_n\rangle$, the states $|\mathcal S^\prime_n\rangle$ are not frustration-free eigenstates of $H$.  Nevertheless, we show below that these states nullify the nearest-neighbor hopping terms in Eq.~\eqref{eq:H-XY} when PBC are imposed, and that a finite density of momentum-$\pi$ bond bimagnons behave effectively like free particles due to destructive interference of local scattering processes.

We begin with the state $|\mathcal S^\prime_1\rangle$, which can be written as
\begin{align}
|\mathcal S^\prime_1\rangle=\frac{1}{\sqrt{L}}\sum^L_{j=1} (-1)^j\, |\,\underset{j-1}{\underbrace{-\cdots}}\,00\underset{L-j-1}{\underbrace{-\cdots}}\,\rangle.
\end{align}
The nearest-neighbor exchange Hamiltonian $H_1$ [Eq.~\eqref{appeq:H1}] acts on this state as
\begin{align}
\label{appeq:H1S'1}
H_1|\mathcal S^\prime_1\rangle
&\propto
\sum^L_{j=1} (-1)^j
\bigg(
|\,\underset{j-2}{\underbrace{-\cdots}}\,0-0\underset{L-j-1}{\underbrace{-\cdots}}\,\rangle
\!+\!
|\,\underset{j-1}{\underbrace{-\cdots}}\,0-0\underset{L-j-2}{\underbrace{-\cdots}}\,\rangle
\!+\!
|\,\underset{j-1}{\underbrace{-\cdots}}+-\underset{L-j-1}{\underbrace{-\cdots}}\,\rangle
\!+\!
|\,\underset{j-1}{\underbrace{-\cdots}}-+\underset{L-j-1}{\underbrace{-\cdots}}\,\rangle
\bigg)\nonumber\\
&=
(1-1)
\sum^L_{j=1} (-1)^j
\bigg(
|\,\underset{j-1}{\underbrace{-\cdots}}\,0-0\underset{L-j-2}{\underbrace{-\cdots}}\,\rangle
\!+\!
|\,\underset{j-1}{\underbrace{-\cdots}}+-\underset{L-j-1}{\underbrace{-\cdots}}\,\rangle
\bigg)=0,
\end{align}
where in going from the first to the second line we shifted the summation indices in the first and last terms in parentheses.  This demonstrates that the bond bimagnons are stable bound states at momentum $\pi$.

Next we consider $|\mathcal S^\prime_2\rangle$, which tells us about pairwise scattering of bond bimagnons.  This state can be written as
\begin{align}
|\mathcal S^\prime_2\rangle
=
\binom{L}{2}^{-\frac{1}{2}}\sum_{j_1\neq j_2} (-1)^{j_1+j_2}\, |(j_1,j_1+1),(j_2,j_2+1)\rangle,
\end{align}
where $|(j,j+1)\rangle$ denotes a bimagnon on the bond $(j,j+1)$.  In considering the action of $H_1$ on $|\mathcal S^\prime_2\rangle$, there are three classes of configurations to be singled out: those for which $|j_2-j_1|\equiv r_{12}=1$, $r_{12}=2$, and $r_{12}=3$.  Configurations with $r_{12}>3$ can simply be handled analogously to Eq.~\eqref{appeq:H1S'1}: at fixed $j_1$, say, the terms $j_2$ and $j_2-1$ enter with opposite signs and interfere destructively. We therefore break up the state $|\mathcal S^\prime_2\rangle$ into disjoint pieces,
\begin{align}
\label{appeq:S'2-decomp}
|\mathcal S^\prime_2\rangle=|\mathcal S^\prime_2\rangle_{\leq 3}+|\mathcal S^\prime_2\rangle_{> 3}=\binom{L}{2}^{-\frac{1}{2}}\[\sum_{|j_2-j_1|\leq 3}+\sum_{|j_2-j_1|> 3} \](-1)^{j_1+j_2}\, |(j_1,j_1+1),(j_2,j_2+1)\rangle,
\end{align}
with
\begin{align}
|\mathcal S^\prime_2\rangle_{\leq 3}\propto \sum_{j_1}\Big(-|\cdots-0+0-\cdots\rangle+|\cdots-0000-\cdots\rangle-|\cdots-00-00-\cdots\rangle\Big).
\end{align}
Next, we compute the action of $H_1$ on the above configurations (keeping track of the signs for later):
\begin{subequations}
\label{appeq:H1S'2leq3}
{\scriptsize
\begin{align}
\label{appeq:H1S'2leq3a}
-H_1|\cdots-0+0-\cdots\rangle
&=
-\bigg(|\cdots-0-+0-\cdots\rangle+|\cdots-+00-\cdots\rangle+|\cdots-00+-\cdots\rangle+|\cdots-0+-0-\cdots\rangle\bigg)\\
\begin{split}
\label{appeq:H1S'2leq3b}
+H_1|\cdots-0000-\cdots\rangle
&=
+\bigg(|\cdots-0-000-\cdots\rangle+|\cdots-000-0-\cdots\rangle+|\cdots-+-00-\cdots\rangle+|\cdots--+00-\cdots\rangle\\
&
+|\cdots-0+-0-\cdots\rangle+|\cdots-0-+0-\cdots\rangle+|\cdots-00+--\cdots\rangle+|\cdots-00-+-\cdots\rangle
\bigg)
\end{split}\\
\label{appeq:H1S'2leq3c}
-H_1|\cdots-00-00-\cdots\rangle
&=-\bigg(
|\cdots-0-0-00-\cdots\rangle
+|\cdots-00-0-0-\cdots\rangle
+|\cdots-0-000-\cdots\rangle
+|\cdots-000-0-\cdots\rangle\nonumber\\
&
+|\cdots-+--00-\cdots\rangle
+|\cdots-+-00-\cdots\rangle
+|\cdots-00-+-\cdots\rangle
+|\cdots-00--+-\cdots\rangle
\bigg).
\end{align}
}%
\end{subequations}
One immediately sees that all terms in Eq.~\eqref{appeq:H1S'2leq3a} are canceled by terms in Eq.~\eqref{appeq:H1S'2leq3b} upon summation over $j_1$ and appropriate shifting of the summation indices.  Similarly, the remaining terms in Eq.~\eqref{appeq:H1S'2leq3b} are canceled by terms in Eq.~\eqref{appeq:H1S'2leq3c}, so that we are left with
\begin{align}
\begin{split}
H_1|\mathcal S^\prime_2\rangle_{\leq 3}
&=
-\sum_{i}
\bigg(
|\cdots-0-0-00-\cdots\rangle
+|\cdots-00-0-0-\cdots\rangle\\
&\qquad\qquad\qquad
+|\cdots-00--+-\cdots\rangle
+|\cdots-+--00-\cdots\rangle
\bigg)
\end{split}
\end{align}
These remaining terms cancel against terms with $d_{12}=4$ in $|\mathcal S^\prime_2\rangle_{> 3}$, which enter with the opposite sign due to the factor $(-1)^{j_1+j_2}$.  The remainder of the terms in $|\mathcal S^\prime_2\rangle_{> 3}$ cancel among themselves as described above Eq.~\eqref{appeq:S'2-decomp}.

The above calculation demonstrates that the pairwise scattering of bond bimagnons interferes destructively when each bimagnon has momentum $\pi$. This property allows for the construction of the many-bond-bimagnon eigenstates $|\mathcal S^\prime_n\rangle$ despite the fact that these eigenstates are not frustration-free like the site bimagnon states $|\mathcal S_n\rangle$. PBC and translation invariance are essential for this construction, as they enable the necessary cancellations to occur.  Moreover, the nearest-neighbor nature of the exchange term $H_1$ is important: longer-range hopping terms including the $J_3$ term used in the main text destroy these eigenstates.\footnote{Since the $J_3$ exchange term was used in the main text to break the hidden SU(2) symmetry of Ref.~\cite{Kitazawa03}, one might wonder whether the states $|\mathcal S^\prime_n\rangle$ are simply consequences of this symmetry.  However, this is not the case because these states do not persist for $D\neq 0$, while the Hamiltonian \eqref{eq:H-XY} retains its hidden SU(2) symmetry for any $D$. Moreover, the existence of the SU(2) symmetry and $|\mathcal S^\prime_n\rangle$ require different boundary conditions.} We note that although we have not strictly proven the existence of the tower states for $n>2$, we have numerically verified their presence up to $L=10$, thus giving strong evidence for the tower based on the scattering mechanism described above. 

Despite their fragility relative to their counterparts $|\mathcal S_n\rangle$ studied in the main text, the states $|\mathcal S^\prime_n\rangle$ present a number of interesting questions for future work:
\begin{enumerate}
\item How does the entanglement entropy of these states scale with $n$ and $L$?
\item Does there exist a set of operators analogous to Eq.~\eqref{eq:su(2)} in the main text that can be used to understand these states?
\item Do these states exhibit off-diagonal long-range order? That is, is there a local order parameter that yields an analogue of Eq.~\eqref{eq:ODLRO} in the main text for the states $|\mathcal S^\prime_n\rangle$?
\item Do these states have any signatures in dynamics?
\item Can analogues of these states be defined for $d>1$?
\end{enumerate}
All of these questions are worthwhile subjects for future work.  Intriguingly, many of these questions are also open for the AKLT bimagnon states studied in Refs.~\cite{Moudgalya18a,Moudgalya18b}; one might speculate that studying these questions for the simpler states $|\mathcal S^\prime_n\rangle$ constructed here could shed some light on the AKLT case as well.

\section{Appendix B: Entanglement of Scar States $|\mathcal S_n\rangle$}

\setcounter{equation}{0}
\renewcommand{\theequation}{B\arabic{equation}}

In this Appendix we compute the entanglement spectrum of the scar states $|\mathcal S_n\rangle$. We also show how to extract the asymptotic expression for the entanglement entropy for the case of half filling, $m_n=0$, which is quoted in Eq.~\eqref{eq:log}.

\subsection{Calculation of Entanglement Spectrum and Entanglement Entropy}

We consider the state $|\mathcal S_n\rangle$ in a system with volume $V$, which we bipartition into regions $A$ and $B$ with volumes $V_A$ and $V_B=V-V_A$.  The entanglement spectrum of the state $|\mathcal S_n\rangle$ is given by the set of eigenvalues $\lambda_k$ ($k=0,\dots,K$) of its reduced density matrix,
\begin{align}
\rho_A(n)=\text{tr}_B |\mathcal S_n\rangle\langle\mathcal S_n|=\sum^{K}_{k=0}\lambda_k\, |\lambda_k\rangle\langle\lambda_k|,
\end{align}
where $|\lambda_k\rangle$ are the associated eigenvectors.  We represent the state $|\mathcal S_n\rangle$ in the local $S^z$ basis as
\begin{subequations}
\begin{align}
|\mathcal S_n\rangle=\binom{V}{n}^{-\frac{1}{2}}\sum_{i_1\neq\dots\neq i_n}\sigma(i_1,\dots,i_n)\, |i_1,\dots,i_n\rangle,
\end{align}
where $i_p=1,\dots,V$ ($p=1,\dots,n$),
\begin{align}
\sigma(i_1,\dots,i_n)=e^{i(\sum^n_{p=1}\bm{r}_p)\cdot\bm{\pi}}
\end{align}
is the alternating phase factor in Eq.~\eqref{eq:Jpm}, and
\begin{align}
|i_1,\dots,i_n\rangle=\bigotimes_j
\begin{cases} 
\left|+\right\rangle & j\in\{i_1,\dots,i_n\}\\
\left|-\right\rangle & \text{otherwise}
\end{cases}
\end{align}
is a product state in the local $S^z$ basis.
\end{subequations}
Because the entanglement spectra of $\rho_A(n)$ and $\rho_A(L-n)$ are related by a ``particle-hole" transformation wherein the local $S^z$-basis states $m_i=\pm 1$ are interchanged, it suffices to consider $n\leq L/2$, which implies that
\begin{align}
K=\text{min}(n,V_A).
\end{align}

To calculate the entanglement spectrum, we bipartition the state $|\mathcal S_n\rangle$ as
\begin{align}
|\mathcal S_n\rangle
=
\sum_{\{S^z_A\},\{S^z_B\}}
M_{\{S^z_A\},\{S^z_B\}}
\left|\{S^z_A\}\right\rangle
\otimes
\left|\{S^z_B\}\right\rangle,
\end{align}
where $\{S^z_{A,B}\}$ denote $S^z$-basis configurations in subsystems $A$ and $B$, respectively.  The entanglement spectrum $\{\lambda_k\}$ is then given by the eigenvalues of $M^\dagger M$ (or, equivalently, $MM^\dagger$).  The form of the matrix $M$ can be read off from the state $|\mathcal S_n\rangle$ when written in the following form:
\begin{align}
|\mathcal S_n\rangle
=
\binom{V}{n}^{-\frac{1}{2}}
\!
\sum^K_{k=0}
\(\sum_{i^A_1\neq\dots\neq i^A_k}
\!\!
\sigma(i^A_1,\dots,i^A_k)
|i^A_1,\dots,i^A_k\rangle\)
\!\!
\otimes
\!\!
\(\sum_{i^B_{k+1}\neq\dots\neq i^B_{n-k}}
\!\!\!\!
\sigma(i^B_{k+1},\dots,i^B_{n-k})
|i^B_{k+1},\dots,i^B_{n-k}\rangle\),
\end{align}
where $i^{A,B}_p=1,\dots,V_{A,B}$ ($p=1,\dots,k$) are defined to reside entirely within subsystems $A$ and $B$, respectively.  One then finds that the matrix $MM^\dagger$ is a symmetric matrix that decomposes into a direct sum of $K$ blocks, each of which has dimension $\binom{V_A}{k}$.  Each block contributes precisely one nonzero eigenvalue, namely
\begin{subequations}
\label{appeq:S_A}
\begin{align}
\label{appeq:lambda_k}
\lambda_{k}=\frac{\binom{V_A}{k}\binom{V_B}{n-k}}{\binom{V}{n}},
\end{align}
and the entanglement entropy is given by
\begin{align}
\label{appeq:S_An}
S_A(n)=-\sum^{K}_{k=0}\lambda_{k}\ln\lambda_{k}.
\end{align}
\end{subequations}
Note that 
\begin{align}
\sum^{K}_{k=0}\lambda_k = \binom{V}{n}^{-1}\sum^{K}_{k=0}\binom{V_A}{k}\binom{V_B}{n-k}=\binom{V}{n}^{-1}\sum^{n}_{k=0}\binom{V_A}{k}\binom{V_B}{n-k}=1,
\end{align}
where the second equality follows from the fact that $\binom{V_A}{k}\equiv0$ for $k>V_A$ (which is necessary only when $n>V_A$ so that $K=V_A$) and the third follows from Vandermonde's identity.  Thus the eigenvalues \eqref{appeq:lambda_k} are properly normalized.

\subsection{Asymptotic Behavior of Entanglement Entropy at Half Filling}

We now show how Eq.~\eqref{eq:log} follows from Eqs.~\eqref{appeq:S_A} for $V_A=V_B=V/2$ and $n=V/2$ in the limit $V\to\infty$.  After substituting $V_A=V_B=n=V/2$ into Eq.~\eqref{appeq:lambda_k}, we apply Stirling's approximation,
\begin{align}
p!\approx \sqrt{2\pi p}\ e^{p\ln p-p},
\end{align}
to obtain
\begin{subequations}
\begin{align}
\lambda_k \approx \frac{e^{V\[s\(\frac{2k}{V}\)-\ln 2\]}}{\sqrt{2\pi V}\ \frac{2k}{V}\(1-\frac{2k}{V}\)},
\end{align}
where we have introduced the Shannon entropy,
\begin{align}
s(x)=-x\ln x-(1-x)\ln(1-x),
\end{align}
\end{subequations}
and where we have assumed that both $k$ and $\frac{V}{2}-k$ are of order $V$ in anticipation of the fact that terms with $k\sim V$ predominate in the sum in Eq.~\eqref{appeq:S_An}.  Next, we define $q=2k/V$ and convert the sum \eqref{appeq:S_An} to an integral over $q$, obtaining
\begin{align}
\begin{split}
S_A\(\frac{V}{2}\)
&\approx
-\frac{V}{2}\int^1_0\mathrm dq\, 
\frac{e^{V\[s\(q\)-\ln 2\]}}{\sqrt{2\pi V}\ q(1-q)}
\left\{V[s(q)-\ln 2]-\ln\[\sqrt{2\pi V}\, q(1-q)\]\right\}\\
&=-\frac{V}{2}\int^1_0\mathrm dq\, 
\frac{1}{\sqrt{2\pi V}\ q(1-q)}\left\{\[\partial_\alpha\, e^{\alpha V\[s\(q\)-\ln 2\]}\]_{\alpha=1}-e^{V\[s\(q\)-\ln 2\]}\ln\[\sqrt{2\pi V}\, q(1-q)\]\right\}.
\end{split}
\end{align}
We proceed to evaluate the integral by saddle point.  We first find the value $q_*$ such that 
\begin{align}
\partial_q\ \alpha V\[s\(q\)-\ln 2\]=\alpha V\, s'(q)=0\implies q_*=\frac{1}{2},\ s(q_*)=\ln 2,
\end{align}
where the prime symbol denotes differentiation with respect to $q$.  Expanding the argument of each exponential around $q_*$ to leading order and noting that $s''(q_*)=-4$, we obtain 
\begin{align}
\label{appeq:S_A-intermediate}
S_A\(\frac{V}{2}\)
&\approx
-\sqrt{\frac{2V}{\pi}}
\int^1_0\mathrm dq\, 
\left\{\[\partial_\alpha\, e^{-2\alpha V\(q-\frac{1}{2}\)^2}\]_{\alpha=1}-\frac{1}{2}e^{-2 V\(q-\frac{1}{2}\)^2}\ln\(\frac{\pi V}{8}\)\right\},
\end{align}
where we have used the fact that the convexity of $s(q)$ renders the integrand sharply peaked around $q_*$.
It remains to evaluate the Gaussian integral
\begin{subequations}
\begin{align}
\int^1_0\mathrm dq\, e^{-2\alpha V\(q-\frac{1}{2}\)^2}=\sqrt{\frac{\pi}{2\alpha V}}
\end{align}
and its derivative,
\begin{align}
\partial_\alpha\int^1_0\mathrm dq\, e^{-2\alpha V\(q-\frac{1}{2}\)^2}=\partial_\alpha\sqrt{\frac{\pi}{2\alpha V}}=-\frac{1}{2}\sqrt{\frac{\pi}{2V}}\ \alpha^{-3/2},
\end{align}
\end{subequations}
which we substitute into Eq.~\eqref{appeq:S_A-intermediate} to obtain
\begin{align}
S_A\(\frac{V}{2}\)
&\approx
-\sqrt{\frac{2V}{\pi}}
\[-\frac{1}{2}\sqrt{\frac{\pi}{2V}}-\frac{1}{2}\sqrt{\frac{\pi}{2V}}\ln\(\frac{\pi V}{8}\)\]=\frac{1}{2}\[\ln\(\frac{\pi V}{8}\)+1\].
\end{align}
This is precisely Eq.~\eqref{eq:log}.

As a consistency check, note that the above large-$V$ analysis maintains the normalization of the eigenvalues $\lambda_k$.  Expanding around the same saddle point, we find
\begin{align}
\sum^{V/2}_{k=0}\lambda_k\underset{V\to\infty}{\longrightarrow}\sqrt{\frac{2V}{\pi}}\int^1_0\mathrm dq\, e^{-2V\(q-\frac{1}{2}\)^2}=1,
\end{align}
as desired.

\section{Appendix C: Relationship to Embedded Hamiltonians}

\setcounter{equation}{0}
\renewcommand{\theequation}{C\arabic{equation}}

The results of this paper have an interesting connection to the method of embedded Hamiltonians developed in Ref.~\cite{Shiraishi17} and applied to quantum many-body scars (QMBS) in Refs.~\cite{Choi18,Ok19,Shiraishi19}.  In this method, one chooses a set of projection operators $P_i$ acting near site $i$ which define the target space
\begin{align}
\mathcal T = \big\{|\psi\rangle \mid P_i\, |\psi\rangle = 0\ \forall\ i\big\}.
\end{align}
Next, one defines a Hamiltonian of the form
\begin{align}
\label{appeq:H-E}
H_\mathrm{E}=\sum_i P_i\, h_i\, P_i + H',
\end{align}
where $h_i$ acts near site $i$ and $[H',P_i]=0$ for all $i$.  Since the first term in Eq.~\eqref{appeq:H-E} annihilates any state in $\mathcal T$, $H_\mathrm{E}$ possesses a number of eigenstates equal to $\text{dim }\mathcal T$  that are also eigenstates of $H'$.  When $H'$ possesses lightly entangled eigenstates that also belong to $\mathcal T$, the Hamiltonian~\eqref{appeq:H-E} hosts strong-ETH-violating eigenstates by construction.

It is natural to suspect that the Hamiltonian $H$ in Eq.~\eqref{eq:H-XY} can be re-expressed in the form~\eqref{appeq:H-E}: the states $|\mathcal S_n\rangle$ are annihilated by the exchange term $H_1$ (as shown explicitly in Sec.~\ref{sec: Appendix A}) and are also eigenstates of the $D$ and $h$ terms, which we can identify with $H'$. We can thus rewrite Eq.~\eqref{eq:H-XY} in the form \eqref{appeq:H-E} if we can recast $H_1$ in terms of appropriate projectors $P_i$ and local Hamiltonian terms $h_i$. In Sec.~\ref{sec:P} we identify a set of such operators, but show that the resulting embedding is \textit{tautological} in the sense that the null spaces of $h_i$ and $P_i$ coincide---in other words, one can simply replace the first term in Eq.~\eqref{appeq:H-E} by $\sum_i h_i=H_1$.  The embedding picture thus does not add anything to our understanding of the states $|\mathcal S_n\rangle$ for the model studied in the main text.  Nevertheless, we show in Sec.~\ref{sec:algebraic} that the projectors $P_i$ have an appealing interpretation in terms of the emergent SU(2) algebra \eqref{eq:su(2)}.  We also show in Sec.~\ref{sec:perturbations} that we can use these projectors to construct a class of perturbations to Eq.~\eqref{eq:H-XY} that preserve the scarred eigenstates $|\mathcal S_n\rangle$.

\subsection{Identifying the Projector}
\label{sec:P}

We now identify the desired operators $P_i$ and $h_i$.\footnote{We thank C.-J.~Lin for bringing to our attention the existence of the operator $P_{i,j}$.  The present derivation and subsequent interpretation of this operator are our own.}
We write the nearest-neighbor exchange Hamiltonian $H_1$ as a sum of local nearest-neighbor bond terms,
\begin{align}
H_1 = \sum_{\langle ij\rangle}h^1_{i,j}, \indent h^1_{i,j}=\frac{1}{2}(S^+_iS^-_j+S^-_iS^+_j).
\end{align}
The local Hilbert space for the two sites $i,j$ is nine-dimensional.
In Sec.~\ref{sec:hardcore} we showed that the null space of $h^1_{i,j}$ is spanned by the three states $|\pm\pm\rangle$ and $|+-\rangle-|-+\rangle$.  We want to define a projection operator $P_{i,j}$ that annihilates these three states and acts trivially on the remaining six.  We claim that the operator
\begin{align}
\label{appeq:P explicit}
P_{i,j}
=
1
-\frac{3}{4}(S^z_i)^2(S^z_j)^2
+\frac{1}{8}\Big[(S^+_i)^2(S^-_j)^2+(S^-_i)^2(S^+_j)^2\Big]
-\frac{1}{4}S^z_iS^z_j
\end{align}
achieves this.  First let us check the action on the six states not in the null space of $h^1_{i,j}$. We have
\begin{align}
P_{i,j}|a\rangle=|a\rangle \indent \text{for} \indent |a\rangle = |00\rangle,\ |0\pm\rangle,\ |\pm0\rangle,
\end{align}
since the non-identity terms in Eq.~\eqref{appeq:P explicit} annihilate any state with a ``0" on site $i$ or $j$.  Furthermore, we have
\begin{align}
\begin{split}
\frac{1}{8}\Big[(S^+_i)^2(S^-_j)^2+(S^-_i)^2(S^+_j)^2\Big]\big(|+-\rangle+|-+\rangle\big) 
&= \frac{1}{2}\big(|+-\rangle+|-+\rangle\big),\\
\Big[1
-\frac{3}{4}(S^z_i)^2(S^z_j)^2
-\frac{1}{4}S^z_iS^z_j\Big]
\big(|+-\rangle+|-+\rangle\big) 
&= \bigg(1-\frac{3}{4}+\frac{1}{4}\bigg)\big(|+-\rangle+|-+\rangle\big)\\
&= \frac{1}{2}\big(|+-\rangle+|-+\rangle\big),
\end{split}
\end{align}
which delivers
\begin{align}
P_{i,j}\big(|+-\rangle+|-+\rangle\big)=\big(|+-\rangle+|-+\rangle\big).
\end{align}
Next, we check the action on the three states in the null space of $h^1_{i,j}$. First, we have
\begin{align}
P_{i,j}|b\rangle=\bigg(1-\frac{3}{4}-\frac{1}{4}\bigg)|b\rangle=0 \indent \text{for} \indent |b\rangle = |\pm\pm\rangle.
\end{align}
We then note that
\begin{align}
\begin{split}
\frac{1}{8}\Big[(S^+_i)^2(S^-_j)^2+(S^-_i)^2(S^+_j)^2\Big]\big(|+-\rangle-|-+\rangle\big) 
&= -\frac{1}{2}\big(|+-\rangle-|-+\rangle\big),\\
\Big[1
-\frac{3}{4}(S^z_i)^2(S^z_j)^2
-\frac{1}{4}S^z_iS^z_j\Big]
\big(|+-\rangle-|-+\rangle\big) 
&= \bigg(1-\frac{3}{4}+\frac{1}{4}\bigg)\big(|+-\rangle-|-+\rangle\big)\\
&= \frac{1}{2}\big(|+-\rangle-|-+\rangle\big),
\end{split}
\end{align}
which delivers
\begin{align}
P_{i,j}\big(|+-\rangle-|-+\rangle\big)=0.
\end{align}
Since $P_{i,j}$ and $h^{1}_{i,j}$ share the same null space, we have
\begin{align}
P_{i,j}h^{1}_{i,j} = h^{1}_{i,j}P_{i,j} = h^{1}_{i,j}.
\end{align}
We may thus trivially rewrite
\begin{align}
H_1=\sum_{\langle ij\rangle}P_{i,j}\, h^{1}_{i,j}\, P_{i,j},
\end{align}
so that Eq.~\eqref{eq:H-XY} becomes Eq.~\eqref{appeq:H-E} with $P_i=P_{i,j}$, $h_i=h^{1}_{i,j}$, and $H'=h\sum_i S^z_i+D\sum_i \left(S^{z}_i\right)^2$.  Note that $[H',P_{i,j}]=0$ because $H'$ does not couple the null space of $P_{i,j}$ and its complement.

\subsection{Algebraic Interpretation}
\label{sec:algebraic}

The projector~\eqref{appeq:P explicit} has an appealing interpretation in terms of the SU(2) algebra~\eqref{eq:su(2)}.  To see it, we define the local \textit{staggered} SU(2) generators
\begin{align}
J^\pm_{i}
=\frac{1}{2}\times
\begin{cases}
-(S^+_i)^2 & i\in \text{sublattice } A\\
+(S^+_i)^2 & i\in \text{sublattice } B
\end{cases}\ ,
\qquad
J^z_i = \frac{1}{2}\, S^z_i,
\end{align}
which satisfy
$\sum_i J^\pm_i = J^\pm$ and $\sum_i J^z_i=J^z$, where the latter operators are the generators in Eq.~\eqref{eq:su(2)}.  In terms of these generators, Eq.~\eqref{appeq:P explicit} becomes
\begin{align}\label{appeq:P-su(2)}
P_{i,j}
=
\frac{3}{4}\bigg(1
-(S^z_i)^2(S^z_j)^2\bigg)
+\bigg(\frac{1}{4}
-\bm J_i\cdot\bm J_j\bigg),
\end{align}
where we have used the fact that sites $i$ and $j$ belong to different sublattices owing to the bipartite nature of the nearest-neighbor exchange.
The first term above vanishes for any configuration with $m_i^2=m_j^2=1$; in particular, it vanishes when acting on the scarred eigenstates $|\mathcal S_n\rangle$.  The second term above projects onto the \emph{local} staggered SU(2) singlet state between sites $i$ and $j$.  Thus, when acting on a state in the $m_i^2=1$ subspace in which the scarred states reside, $P_{i,j}$ annihilates any state in the \emph{global} SU(2) representation with maximal spin $j=V/2$ (a state with maximal total spin cannot contain any local singlets).  Conversely, $P_{i,j}$ acts as the identity on any state with $m_{i,j}=0$ for one or both of $i$ and $j$ due to the fact that $\bm J_i\cdot\bm J_j$ annihilates such states.

The above rewriting of $P_{i,j}$ thus makes explicit the fact that the states $|\mathcal S_n\rangle$ form components of a $(V+1)$-state ``macrospin."  In this way, the model studied in the main text resembles the toy model proposed in Ref.~\cite{Choi18} that has perfect QMBS. However, it is worth noting that while the projectors $P_{i,j}$ are superfluous for the model studied in the main text (as mentioned elsewhere in this Appendix), they are not for the model proposed in Ref.~\cite{Choi18}. There, the analogous projectors themselves enforce the presence of scarred eigenstates and no structure for the Hamiltonian terms analogous to $h_i$ in Eq.~\eqref{appeq:H-E} is assumed, while in the example presented here these states exist regardless of the presence or absence of the projectors $P_{i,j}$ due to the structure of the local exchange terms $h^1_{i,j}$.

\subsection{A Class of Perturbations Preserving the Scar States}
\label{sec:perturbations}

Despite the fact that the projectors defined in Eq.~\eqref{appeq:P explicit} are superfluous for the XY model defined Eq.~\eqref{eq:H-XY}, they are useful in identifying a class of perturbations to Eq.~\eqref{eq:H-XY} that preserve the scarred eigenstates $|\mathcal S_n\rangle$.  This more general class of Hamiltonians takes the form
\begin{align}
\label{appeq:general}
H = H_1 + H' + \sum_{i\in A,j\in B}P_{i,j}\, h'_{i,j}\, P_{i,j},
\end{align}
where $h'_{i,j}$ is \emph{any} Hamiltonian coupling sites $i$ and $j$ and $H'$ is \emph{any} Hamiltonian that commutes with all the $P_{i,j}$ appearing in the sum.  Note that the sum above is not restricted to nearest-neighbors, to emphasize that couplings of arbitrary range are allowed as long as the two sites being connected belong to different sublattices.  The projector $P_{i,j}$ annihilates any state in the spin-$V/2$ representation of the algebra~\eqref{eq:su(2)} because such states cannot contain staggered-SU(2) singlets between \emph{any} two spins on different sublattices.  Eq.~\eqref{appeq:general} suggests that the XY Hamiltonian $H_1$ can be viewed as a ``parent Hamiltonian" for a class of embedded-Hamiltonian models.

\section{Appendix D: Spin-$S$ Models with Exact Scarred States}
\label{sec:spin-s}

\setcounter{equation}{0}
\renewcommand{\theequation}{D\arabic{equation}}

In this Appendix we generalize the spin-1 XY model to achieve exact QMBS for microscopic degrees of freedom $\bm S_i$ with arbitrary spin $S$ [i.e., $\bm S_i\cdot\bm S_i=S(S+1)$]. The generators of the emergent SU(2) algebra \eqref{eq:su(2)} are
\begin{subequations}
\label{appeq:su(2)-s}
\begin{align}\label{appeq:su(2)-s-a}
J^{\pm}&=\sum_i J^\pm_i=\frac{1}{(2S)!}\sum_{i}e^{i\boldsymbol{r}_i\cdot\boldsymbol{\pi}}\left(S^\pm_i\right)^{2S},\\
J^z&=\sum_i J^z_i=\frac{1}{2\left((2S)!\right)^2}\sum_i \left[(1+S-S^z_i)_{2S}(1-S+S^z_i)_{2S}-(1+S+S^z_i)_{2S}(1-S-S^z_i)_{2S}\right],
\end{align}
where we have used the notation
\begin{align}\label{appeq:su(2)-s-b}
(x)_{q}=x(x+1)\dots(x+q-1)
\end{align}
\end{subequations}
for the Pochhammer symbol (or ``rising factorial").
Despite its complicated appearance, $J^z_i$ has a simple action on local $z$-basis states, namely
$J^z_i|m_i=\pm S\rangle=\pm(1/2)|m_i=\pm S\rangle$ and $J^z_i|m_i\neq\pm S\rangle=0$.  When every site is in the $\pm S$ state, the spectrum of $J^z$ is that of a spin-$V/2$ degree of freedom, i.e., $-V/2,-V/2+1,\dots,V/2$. When all but $p$ sites are in the $\pm S$ state, the ``total spin" is effectively lowered by $p/2$, so that $J^z$ has the spectrum $-(V-p)/2,\dots,(V-p)/2$.

For $n=0,\dots, V$ we define the states
\begin{align}\label{appeq:scar-s}
|\mathcal{S}^{(S)}_n\rangle=\mathcal N(n)\left(J^+\right)^n|\Omega\rangle,
\end{align}
where $|\Omega\rangle=\bigotimes_i|m_i=-S\rangle$ is the all-down state and $\mathcal N(n)$ is defined below Eq.~\eqref{eq:scar} in the main text.  States with different $n$ are distinguished by their $J_z$ eigenvalues
\begin{align}\label{appeq:m-s}
m^{(S)}_{n}=n-V/2.
\end{align}
Combining Eqs.~\eqref{appeq:su(2)-s}--\eqref{appeq:m-s}, one verifies that the states $|\mathcal{S}^{(S)}_n\rangle$ form a spin-$V/2$ representation of Eq.~\eqref{eq:su(2)}, i.e.
\begin{align}
\bm J\cdot \bm J\,|\mathcal S^{(S)}_n\rangle=\frac{V}{2}\left(\frac{V}{2}+1\right)|\mathcal S^{(S)}_n\rangle.
\end{align}

We find that the states $|\mathcal S^{(S)}_n\rangle$ are annihilated by the following family of Hamiltonians,
\begin{equation}\label{appeq:H-s}
H^{(S)}_p(i,j)=\left(S^+_i\right)^{2S-p}\left(S^-_j\right)^{2S-p}+\left(\frac{(2S-p)!}{p!}\right)^2 \left(S^+_i\right)^{p}\left(S^-_j\right)^{p}+{\rm H.c.},
\end{equation}
where the sites $i,j$ belong to opposite sublattices and $p=1,\dots,S$ for $S\in\mathbbm Z$ while $p=1,\dots,S-\frac{1}{2}$ for $S\in\mathbbm Z+1/2$. We note that for $S=1$ only $p=1$ is allowed and this is exactly the XY Hamiltonian $H_1$ (up to rescaling). For any $S,p$ one may add the total field $h\sum_i S^z_i$ and anisotropy $D\sum_i \left(S^z_i\right)^2$ to the Hamiltonian, giving the scarred states the energy spectrum $E^{(S)}_n=2S\, h(n-V/2)+S^2DV$.

The models defined by Eqs.~\eqref{appeq:su(2)-s}-\eqref{appeq:H-s} admit a trivial embedding structure similar to the one discussed in Sec.~\ref{sec:algebraic}. In this case, Eq.~\eqref{appeq:P-su(2)} is modified as
\begin{align}
P^{(S)}_{i,j}
=
\frac{3}{4}\bigg[1
-\big(2J^z_i\big)^2\, \big(2J^z_j\big)^2\bigg]
+\bigg(\frac{1}{4}
-\bm J_i\cdot\bm J_j\bigg),
\end{align}
where as before the first term is necessary to ensure that $P^{(S)}_{i,j}|m_{i}m_{j}\rangle=|m_{i}m_{j}\rangle$ when one or both of $m_{i,j}\neq\pm S$ and the second term projects onto the singlet of the local staggered SU(2) algebra corresponding to Eqs.~\eqref{appeq:su(2)-s} when $m_{i,j}=\pm S$.

\end{widetext}

\end{document}